\documentclass[letterpaper, 10 pt, conference]{ieeeconf}  

\IEEEoverridecommandlockouts                              

\usepackage[pdftex]{graphicx}
\usepackage{threeparttable}
\usepackage{url}
\usepackage{times} 
\usepackage{multirow}

\title{\LARGE \bf Inter--stimulus Interval Study for the Tactile Point--pressure Brain--computer Interface}

\author{Kensuke Shimizu$^{1,*}$, Shoji Makino$^{1}$, and Tomasz M. Rutkowski$^{1,2,3,*}$
\thanks{$^{1}$Kensuke Shimizu, Shoji Makino and Tomasz M. Rutkowski are with Life Science Center of TARA and Department of Computer Science,
        University of Tsukuba, 1-1-1 Tennodai Tsukuba Ibaraki, Japan. {\tt\small tomek@bci-lab.info} \qquad {\tt\small http://bci-lab.info/}}
\thanks{$^{2}$Tomasz M. Rutkowski is also with RIKEN Brain Science Institute, Wako-shi, Japan.}
\thanks{$^{3}$Tomasz M. Rutkowski is the corresponding author.}\thanks{$^{*}$ Kensuke Shimizu and Tomasz M. Rutkowski were supported in part by YAMAHA Corporation.}     
        }

\begin{document}

\maketitle
\thispagestyle{empty}
\pagestyle{empty}

\begin{abstract}
The paper presents a study of an inter--stimulus interval (ISI) influence on a tactile point--pressure stimulus--based brain--computer interface's (tpBCI) classification accuracy. A novel tactile pressure generating tpBCI stimulator is also discussed, which is based on a three--by--three pins' matrix prototype. The six pin--linear patterns are presented to the user's palm during the online tpBCI experiments in an oddball style paradigm allowing for ``the aha--responses'' elucidation, within the event related potential (ERP). A subsequent classification accuracies' comparison is discussed based on two ISI settings in an online tpBCI application. A research hypothesis of classification accuracies' non--significant differences with various ISIs is confirmed based on the two settings of 120~ms and 300~ms, as well as with various numbers of ERP response averaging scenarios.
\end{abstract}

\section{INTRODUCTION}

A brain--computer interface (BCI) is a technology the uses only the central nervous system signals (brainwaves) of paralyzed or locked--in syndrome (LIS) patients~\cite{lis139review1986} to create a new communication channel with others or to control external devices without depending on any muscle activity~\cite{bciBOOKwolpaw,youtubeVRandBCI_1}. 
The BCI technology has provided a support already to patients' life improvement who suffer from severe paralysis due to diseases like an amyotrophic lateral sclerosis (ALS)~\cite{bciBOOKwolpaw}. The contemporary BCI applications rely mostly on a visual mode, which generates the most reliable event related potentials (ERP) so far~\cite{MoonJeongBCImeeting2013}. However, many LIS patients
in the advanced disease stages often lose their ability to control reliably even their eyeball focusing or movement abilities~\cite{lis139review1986,tomekJNM2014}, and therefore they need the alternative options for BCI--enabled communication.

The successful alternative options have been developed recently to utilize spatial
auditory~\cite{MoonJeongBCImeeting2013,iwpash2009tomek} or tactile (somatosensory) modes~\cite{tomekJNM2014,HiromuBCImeeting2013}.
Meanwhile, the tactile BCI seems to offer the superior communication options in comparison with the contemporary visual and auditory modes in case of LIS patients~\cite{lis139review1986,JNEtactileBCI2012}. 

We present results of a followup study with an extended number to ten subjects comparing to our previous reports~\cite{tpBCIscis2014kensuke,vrBCIscis2014bertrand}. We test the tactile point--pressure stimulus--based brain--computer
interface (tpBCI) using stimuli generated by a matrix of small solenoids (see Figure~\ref{fig:position}). The tpBCI device can generate various patterns to be applied to possibly different body areas, and therefore it could be adopted to many patient symptoms. The presented approach allows for faster and more precise delivery of tactile pressure stimuli comparing to the previously proposed vibrotactile stimulator--based approaches~\cite{tomekJNM2014,HiromuBCImeeting2013} and it is not limited to finger tips only~\cite{JNEtactileBCI2012}.

The goal of this study is to compare and test the performance (BCI classification accuracy) of the novel tpBCI paradigm~\cite{tpBCIscis2014kensuke,vrBCIscis2014bertrand} in function of two inter--stimulus interval (ISI) settings. Namely the ISI equal to $300$~ms (a very easy for the users) and $120$~ms (a harder case due to fast repetition of the presented pin--pressure patterns and larger overlap of the brainwave ERPs).

Results obtained with ten healthy users' offline analysis with stepwise linear discriminant analysis (SWLDA) classifier~\cite{krusienski2006} and five different averaging settings are analyzed for statistical siginficance.

From now on the paper is organized as follows. In the next section we present methods used and developed in order to capture, process and classify the brainwave response in application to the proposed tpBCI. Offline EEG analysis results together with conclusions summarize the paper.

\section{METHODS}

\subsection{Tactile Pin--pressure Stimulus Device}

\begin{figure}[t]
	\vspace{0.3cm}
	\begin{center}
  	\includegraphics[width=\linewidth]{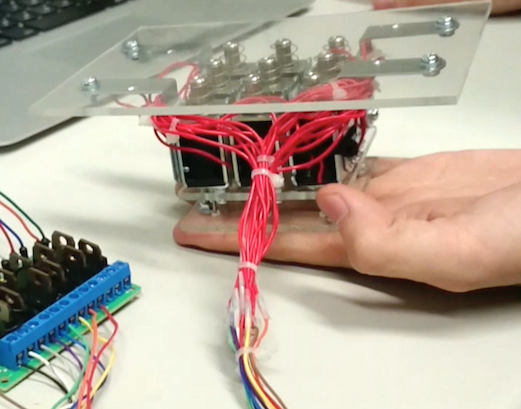}
 	\end{center}
 	\caption{The tactile pressure generator composed of nine solenoids placed on the user's dominant plastic glove covered hand. The solenoids created six linear pressure patterns.}\label{fig:position}
\end{figure}

The tactile stimuli were delivered as light pin--pressure patterns generated by a portable computer with a program developed by our team on the ARDUINO UNO micro--controller board managed from our visual programming application designed in MAX~6~\cite{maxMSP}. Each tactile stimulus pattern was generated via the tactile pressure device composed of nine solenoids arranged in the $3\times3$ matrix as depicted in Figure~\ref{fig:position}. The binary outputs from the ARDUINO UNO micro--controller board were amplified by a multichannel amplifier (battery driven for user's safety) developed also by our team.
There were six linear pattens of tactile pressure stimuli delivered in
random order to the user fingers (for details please refer to~\cite{tpBCIscis2014kensuke}) in order to elicit P300 brainwave responses in an oddball style paradigm~\cite{bciBOOKwolpaw}.

The solenoids generated quick, $100$~ms long, light pin--pressures with inter--stimulus intervals (ISI) of $120$~ms and $300$~ms in two settings as explained in Table~\ref{table:EEGexperimentCondition} with experimental condition details.  The two different ISI setups were tested in order to evaluate a possible impact of the fast stimulation repetitions on user's BCI performance (the resulting BCI accuracy) in five different averaging scenarios. 

\subsection{EEG Experiments} 

\begin{figure}[b]
	\begin{center}
	\includegraphics[width=\linewidth]{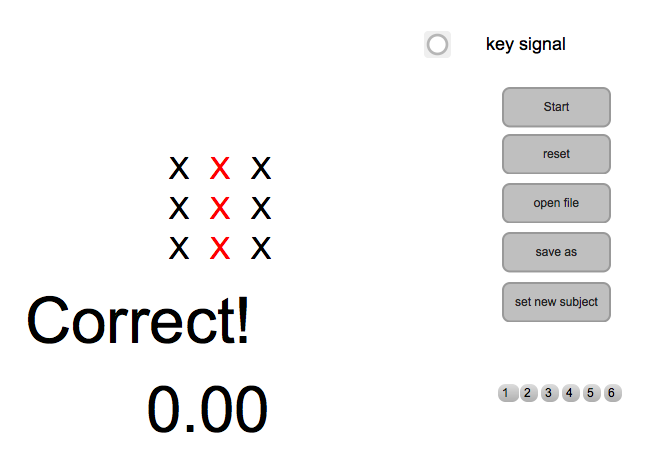}
	\end{center}
	\caption{The visual instruction screen presented to the users during the experiments programmed in MAX~6~\cite{maxMSP}. The red $\times$ symbols inform about the pattern shape to be attended by a user in each SWLDA training experimental run.}\label{fig:max1}
\end{figure}

During the EEG experiments, the users placed the tactile stimulus generator on their  dominant hand's fingers with a plastic glove to prevent any spurious electric current interferences possibly originating from the device. The users responded mentally by confirming/counting only to the instructed patterns while ignoring the others. The training instructions were presented visually by means of the MAX~6~\cite{maxMSP} programmed display designed by our team as depicted in Figure~\ref{fig:max1}.

In the online BCI experiments (similarly as in our previous study~\cite{tpBCIscis2014kensuke}) the EEG signals were captured with a portable EEG
amplifier system g.USBamp from g.tec Medical Instruments,
Austria. Eight active wet EEG electrodes were used to capture brainwaves with attentional modulation elucidated, within the event related potentials (ERPs),   the so-called ``aha-'' or P300-responses~\cite{bciBOOKwolpaw}. The
EEG electrodes were attached to the head locations
\emph{Cz, Cpz, P3, P4, C3, C4, CP5}, and \emph{CP6} as in $10/10$
intentional system~\cite{Jurcak20071600}. A reference
electrode was attached to a left earlobe and a ground electrode on the
forehead at \emph{FPz} position respectively. The experimental details are summarized in Table~\ref{table:EEGexperimentCondition}.

The users were requested to limit their eye--blinks and body muscle movements to avoid electromagnetic and electromyographic interferences. The unavoidable eye--blinks were identified within the ERPs and removed with a threshold setting of $80~\mu$V. All EEG experiments were conducted in the Life Science Center of TARA, University of Tsukuba, Japan.
The details of the experimental procedures and the research targets of the tpBCI paradigm were explained in detail to the seven human users, who agreed voluntarily to participate in the study.
The electroencephalogram (EEG) tpBCI experiments were conducted in accordance with \emph{The World Medical Association Declaration of Helsinki - Ethical Principles for Medical Research Involving Human Subjects}. 
The experimental procedures were approved and designed in agreement with the ethical committee guidelines of the Faculty of Engineering, Information and Systems at University of Tsukuba, Tsukuba, Japan (experimental permission no.~$2013R7$). 

Ten subjects participated in two studies with various ISI settings for tBCI's stimulus speed and difficulty evaluation. The average age of the users was of $24.8$ years old (standard deviation of $3.8$ years old; ten men).
\begin{table}[t]
	\vspace{0.2cm}
    \caption{Conditions and details of the tpBCI EEG experiment}
    \label{table:EEGexperimentCondition}
    \renewcommand{\arraystretch}{1.3}
    \centering
	\begin{tabular}{|l|l|}
	\hline 
	\bf Condition                 	& \bf Detail \\
	\hline \hline
	Number of users            		& $10$ \\
	Tactile stimulus length     	& $100$~ms \\
	Inter-stimulus-interval (ISI)  	& $120$ and $300$~ms \\
	EEG recording system            & g.USBamp active wet electrodes \\
	Number of the EEG channels      & $8$ \\
	EEG electrode positions         & \textsf{Cz, Cpz, P3, P4, C3, C4,} \\
									& \textsf{CP5}, and \textsf{CP6} \\
	Reference electrode            	& Behind the user's left earlobe \\
	Ground electrode                & On the forehead(\textsf{FPz})\\
	Stimulus generator              & $3 \times 3$ pressure pins matrix \\
	Number of trials for each user & $5$ \\
	\hline
	\end{tabular}
\end{table}
\begin{figure}[t]
	\vspace{0.2cm}
	\begin{center}
  	\includegraphics[width=\linewidth]{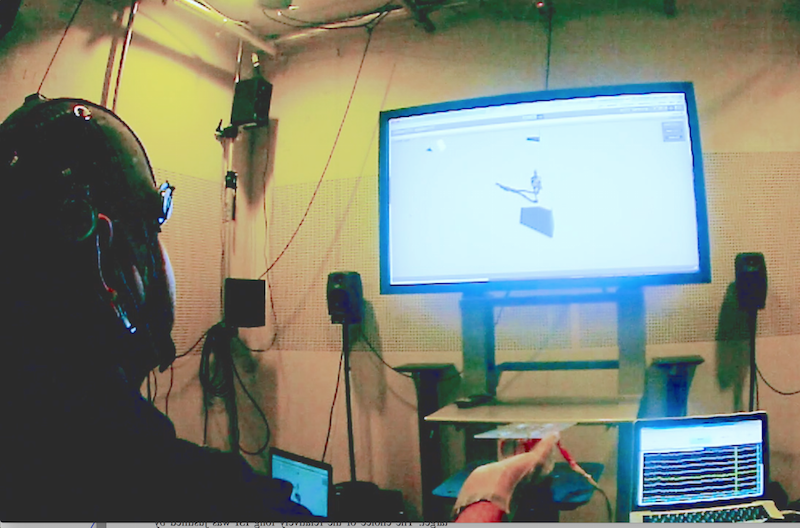}
 	\end{center}
 	\caption{A screenshot from a demo video available online~\cite{youtubeVRandBCI_1} presenting single trial based online tpBCI application to a computer game--based real--time virtual agent walking control (see the last part of the online video).}\label{fig:demo}
\end{figure}

\subsection{EEG Signals Processing and Classification} 

\begin{figure}[t]
	\vspace{0.2cm}
	\begin{center}
	\includegraphics[height=11.5cm]{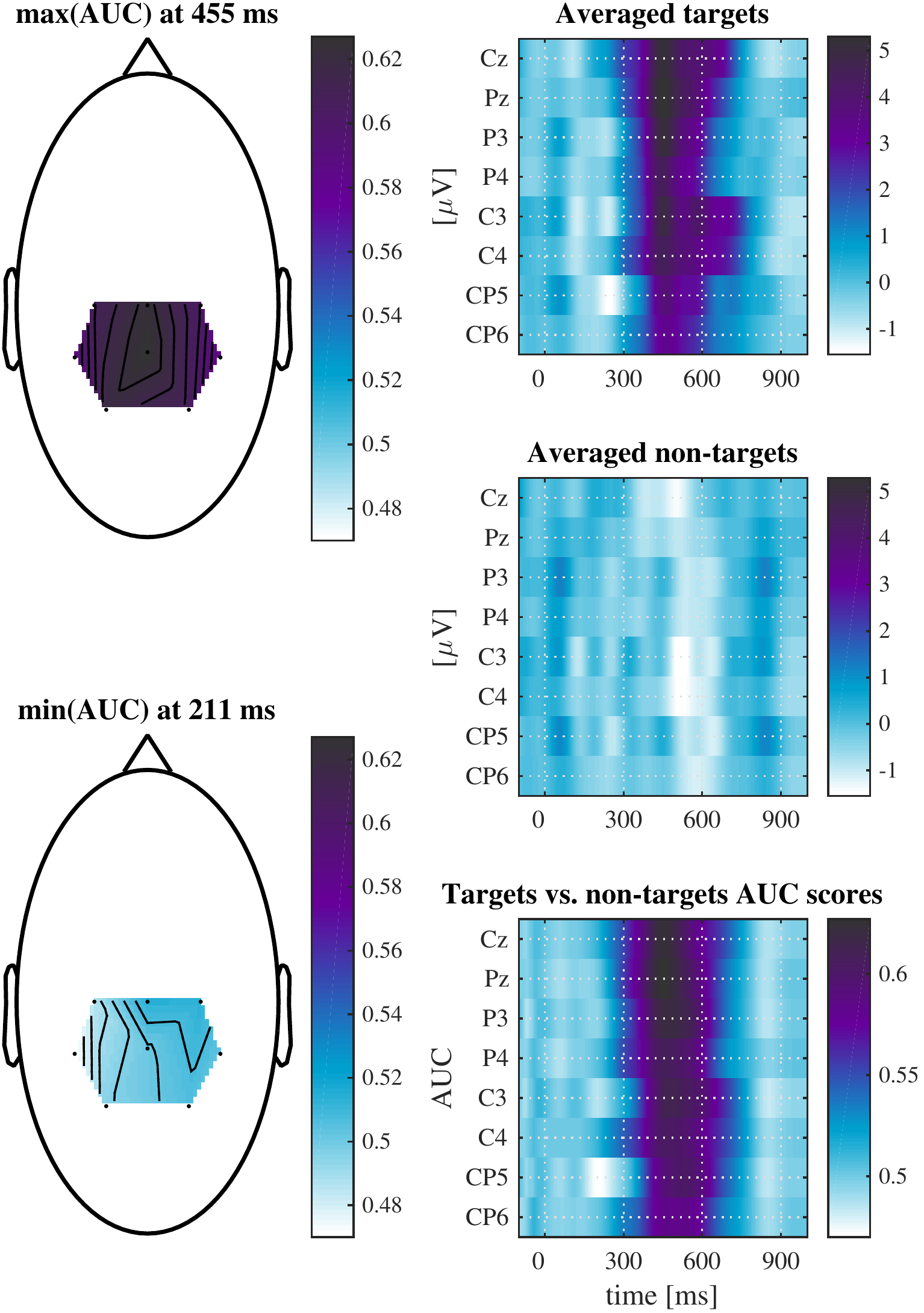}
	\end{center}
	\caption{A summary of the grand mean averaged ERPs from experiments with $ISI=300$~ms presented in left panels. The very clear P300 target responses are depicted with white shades. The right bottom panel presents an area under the curve (AUC) discriminability analysis. The left head topographical maps present the electrode locations together with maximum and minimum AUC scores.}\label{fig:ERP300}
\end{figure}

\begin{figure}[t]
	\vspace{0.2cm}
	\begin{center}
	\includegraphics[height=11.5cm]{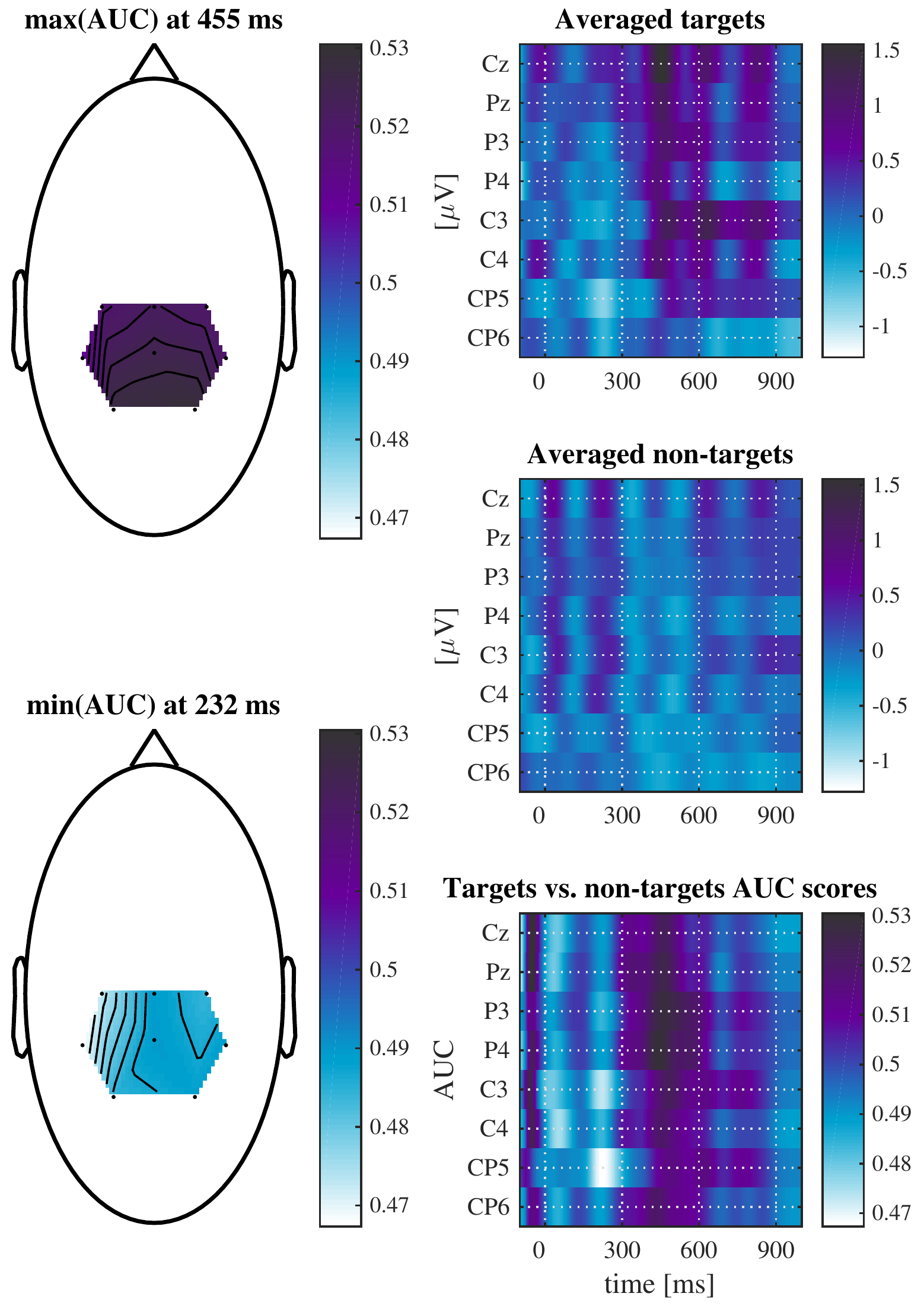}
	\end{center}
	\caption{A summary of the grand mean averaged ERPs from experiments with $ISI=120$~ms presented in left panels. The very clear P300 target responses are depicted with white shades. The right bottom panel presents an area under the curve (AUC) discriminability analysis. The left head topographical maps present the electrode locations together with maximum and minimum AUC scores.}\label{fig:ERP120}
\end{figure}

The EEG signals were recorded and preprocessed online by an in--house extended  BCI2000 application~\cite{bci2000book} and segmented (``epoched'') as features drawn from ERP intervals of $0 \sim 900$~ms. A common average reference (CAR) filter were also applied to the segmented signals.
The sampling rate was set to $512$~Hz, the high pass filter at $0.1$~Hz, and
the low pass filter at $40$~Hz. The ISI were of $120$~ms or $300$~ms in two different experimental runs, and each tactile pressure stimulus duration was of $100$~ms.
Each user performed two sessions of selecting the six patterns (a spelling of a sequence of six digits associated with each tactile pressure pattern).
Each target was presented five times in a random series with the
remaining non--targets. We performed offline analysis of the collected online EEG datasets in order to test a possible influence of the two ISI settings on the tpBCI accuracy (compare ERP results in Figures~\ref{fig:ERP300}~and~\ref{fig:ERP120} depicting an impact of faster ISI on the ERP shapes and areas under the curve (AUC) based discriminative latencies). The stepwise linear discriminant analysis (SWLDA)~\cite{krusienski2006} classifier was applied next, with features drawn from the $0\sim900$~ms ERP interval, with removal of the least significant input features, having $p > 0.15$, and with the final discriminant function restricted to contain a maximum of $60$ features.

A screenshot from an online virtual reality application~\cite{vrBCIscis2014bertrand} of the discussed paradigm demonstrating single trial--based tpBCI operation is presented in Figure~\ref{fig:demo}. A demonstration video is also available online at~\cite{youtubeVRandBCI_1}.

\section{RESULTS}

The results of the SWLDA in comparison of the two ISI settings have been summarized in Table~\ref{tab:results}. The results were slightly better for the longer ISI of $300$~ms, yet this observation was not confirmed with a Wilcoxon rank sum test of a statistical significance for equal medians (the distributions were not normal).

We concluded that the analysis of EEG data obtained from the ten healthy subjects did not confirmed significantly that shorter ISI, causing a larger experimental difficulty, would sabotage the tpBCI accuracy as summarized in Table~\ref{tab:results}.

\section{CONCLUSIONS}

The aim of this study was to test the ISI variability impact on the BCI classification accuracy results in various brainwave averaging scenarios of the proposed tactile pin--pressure BCI paradigm. The offline results obtained with the SWLDA classifier did not show significant differences, although the median tpBCI classification results for ISI of $300$~ms seemed to be higher.

The results are very promising for future online applications with patients suffering from LIS allowing for speeding up the BCI stimuli presentation without significant classification drop danger.

\begin{table}[!t]
	\vspace{0.2cm}
	\begin{center}
	\begin{threeparttable}
	\caption{Single to five trials based tpBCI median accuracy and first quartile ($25^{th}$--percentile) results, without statistical significances at a level of $p\gg0.05$ as tested with the Wilcoxon rank sum test for equal medians}\label{tab:results}
\begin{tabular}{|c|c|c|}
\hline
\multicolumn{3}{|c|}{Mean tpBCI accuracy and standard deviation results of ten subjects} \\
\multicolumn{3}{|c|}{from experiments with ISI $=300$ms}\\
\hline \hline
\multirow{2}{*}{Number of averages} & \multirow{2}{*}{Median accuracy} 	& Data first quartile \\
									&									& ($25^{th}$--percentile) \\
\hline
$1$			& $41.5\%$	&	$21.0\%$ \\
$2$			& $33.0\%$	&	$21.0\%$ \\
$3$			& $50.0\%$	&	$33.0\%$ \\
$4$			& $50.0\%$	&	$33.0\%$ \\
$5$			& $66.5\%$	&	$37.3\%$ \\
\hline \hline
\multicolumn{3}{|c|}{Mean tpBCI accuracy and standard deviation results of ten subjects} \\
\multicolumn{3}{|c|}{from experiments with ISI $=120$ms}\\
\hline \hline
\multirow{2}{*}{Number of averages} & \multirow{2}{*}{Median accuracy} 	& Data first quartile \\
									&									& ($25^{th}$--percentile) \\
\hline
$1$			& $17.0\%$	&	$~9.0\%$\\
$2$			& $33.0\%$	&	$17.0\%$ \\
$3$			& $33.0\%$	&	$25.0\%$ \\
$4$			& $33.0\%$	&	$25.0\%$ \\
$5$			& $33.0\%$	&	$25.0\%$ \\
\hline
\end{tabular}
\end{threeparttable}
\end{center}
\end{table}

The approach presented shall help, if not to reach the goal, to get closer to our objective of the more user friendly tactile BCI design, which would fit those users with bad, or disabled, vision and hearing. Thus, we can expect that patients suffering from LIS will be able to use the appropriate BCI interfaces, according to their intact sensory modalities, more efficiently and comfortably to restore their basic communication needs.

Still there remains a long path to go before providing the effective and comfortable tactile BCI end--user solution, yet our research has progressed toward this goal as presented in this paper.



\section*{ACKNOWLEDGMENTS}

This work was supported by the Strategic Information and Communications R\&D Promotion Program (SCOPE) no. 121803027 of The Ministry of Internal Affairs and Communication in Japan, YAMAHA Corporation research grant, and by KAKENHI, the Japan Society for the Promotion of Science, grant no. 24243062.



\addtolength{\textheight}{-12cm} 

\end{document}